\documentclass[aps,nofootinbib,12pt]{revtex4-2}
\usepackage{graphicx,color}
\usepackage{relsize}
\usepackage{appendix}
\usepackage{latexsym,amsmath,amssymb,graphicx,booktabs}
\usepackage{hyperref}
\usepackage{cleveref}
\usepackage{upgreek}

\newcommand{\unit}{\hat{1\!\!1}}

\graphicspath{ {./} }
\begin{document}

\title{\bf\Large Quantum origin of dark energy and the Hubble tension}

\author{\bf\large Enis Belgacem}
\affiliation{Institute for Theoretical Physics and EMME$\Phi$, 
Faculty of Science, Utrecht University, Princetonplein 5, 3584 CC Utrecht, The Netherlands}

\author{\bf \large Tomislav Prokopec}
\affiliation{Institute for Theoretical Physics and EMME$\Phi$, Faculty of Science, Utrecht University, Princetonplein 5, 3584 CC Utrecht, The Netherlands}

\begin{abstract}

\noindent Local measurements of the Hubble parameter obtained from the distance ladder at low redshift are in tension with global values inferred from cosmological standard rulers. A key role in the tension is played by the assumptions on the cosmological history, in particular on the origin of dark energy.
Here we consider a scenario where dark energy originates from the amplification of quantum fluctuations of a light field in inflation.
We show that spatial correlations inherited from inflationary quantum fluctuations can reduce the Hubble tension down to one standard deviation, thus relieving the problem with respect to the standard cosmological model.
Upcoming missions, like Euclid, will be able to test the predictions of models in this class.

\end{abstract}

\maketitle

\section{Introduction}
\label{sect:intro}

The standard cosmological model, known as $\Lambda$CDM, provides a description of the Universe in agreement with a large number of observations, including the cosmic microwave background (CMB), large-scale structure data from galaxy surveys and distance measurements of type-Ia supernovae (SNeIa). However, the values of some of the parameters in the model show tensions when comparing the results obtained by probing different scales. This is particularly relevant for the Hubble parameter, for which the tension has now overcome 4$\sigma$.

The Hubble tension within $\Lambda$CDM can be described as an incompatibility between $H_0$ measurements from cosmological probes at early times and values deduced from distance measurements at local scales. For simplicity, we will refer to these two classes of observations as global and local measurements, respectively. Global measurements include studies of CMB anisotropies by the {\it Planck} mission~\cite{Planck:2018vyg} and observations of the sound horizon scale imprinted by baryon acoustic oscillations (BAO) on the galaxy correlation function (see \cite{Addison:2017fdm,Schoneberg:2019wmt} for estimation of $H_0$ from BAO with a big bang nucleosynthesis prior on baryon density). It is also possible to study the variation in time of the Hubble parameter up to redshift $z\sim 2$ using the so-called cosmic chronometers~\cite{Jimenez:2001gg,Haridasu:2018gqm}, which are galaxies with known time evolution. Local measurements comprehend distances of SNeIa calibrated with Cepheids by the SH0ES collaboration~\cite{Riess:2019cxk} and time delays of multiple images of strongly lensed quasars studied by H0LiCOW~\cite{Wong:2019kwg}. Among other local probes which have been considered, one can mention calibrating SNeIa with the tip of the red giant branch (TRGB) (finding a lower $H_0$ compared to Cepheids calibration, see~\cite{Freedman:2020dne}) or with Mira variables~\cite{Huang:2018dbn}, measures of distances from surface brightness fluctuations~\cite{Blakeslee:2021rqi} and observations of  galaxies hosting megamasers~\cite{Pesce:2020xfe}. The Hubble parameter estimated from {\it Planck}~\cite{Planck:2018vyg} is $H_0=(67.4\pm 0.5)~{\rm km/s~Mpc^{-1}}$ while the SH0ES collaboration~\cite{Riess:2019cxk} finds $H_0=(74.03\pm 1.42)~{\rm km/s~Mpc^{-1}}$, thus resulting in a tension of more than $4\sigma$.

Several ideas have been proposed in order to reconcile these two values. A possible approach consists in introducing new physics such that the Hubble parameter inferred from CMB is larger. Such modifications can affect the cosmological evolution at late times or even the early Universe. As an example of late-time proposal, it has been shown that a phantom dark energy equation of state shifts the Hubble parameter predicted from CMB towards higher values~\cite{DiValentino:2016hlg}. Other studies consider possible late-time solutions coming from interacting dark energy~\cite{DiValentino:2019ffd}, metastable dark energy~\cite{Li:2019san} or decaying dark matter~\cite{Vattis:2019efj}. Indeed these scenarios give higher values of $H_0$ when fitting CMB data only, but in general the other independent $H_0$ measurements coming from BAO and cosmic chronometers leave little room for such late-times modifications, see~\cite{Feeney:2018mkj,Lemos:2018smw}. As for early-Universe proposals, an example is early dark energy \cite{Karwal:2016vyq,Poulin:2018cxd} where a scalar field dilutes faster than matter starting from an initial state frozen by Hubble friction. Apart from fine-tuning difficulties ({\it e.g.} in the required potential of the scalar field, in its initial conditions or in making early dark energy appear and dilute at the right times), early dark energy still has troubles related to the $\sigma_8$ tension between large-scale structures and CMB~\cite{Ivanov:2020ril}. Further early-Universe proposals try to reduce the sound speed of the primordial photon-baryon plasma before recombination or attempt to make recombination occur earlier~\cite{Jedamzik:2020krr}. While all these ideas deal with changing the prediction of $H_0$ from CMB, other approaches focus instead on local measurements by studying how the SNeIa calibration could be affected by new physics, like chameleon dark energy~\cite{Cai:2021wgv} or late-time changes in the effective Newton constant~\cite{Marra:2021fvf}.
\\

\noindent As an alternative line of thought, in this paper we want to explore the scenario where late-time dark energy originates from quantum fluctuations in the early Universe and we start a study of the consequences expected on the spatial correlations of dark energy today. The analysis will allow us to make quite general predictions on the relief of the Hubble tension within this framework.
In such a scenario the Hubble parameter is a stochastic variable and the stochasticity source comes not only from (dark) matter fluctuations due to classical inhomogeneities, but also from the intrinsic quantum nature of dark energy.
At time $t$ and for every comoving coordinate $\vec{x}$ the effect of quantum dark energy is such that the expansion rate at that point is an operator satisfying the first Friedmann equation
\begin{equation}
\label{eq:pointH2general}
3M_P^2 \hat{H}^2(t,\vec{x})=\rho_{C}(t)\unit+\hat{\rho}_{Q}(t,\vec{x})\,,
\end{equation}
where $M_P\equiv\left(8\pi G\right)^{-1/2}$ is the reduced Planck mass, $\rho_{C}(t)$ is the classical contribution to energy density, $\unit$ is the identity operator and $\hat{\rho}_{Q}(t,\vec{x})$ is the quantum energy density operator\footnote{In general, only a part of the full quantum energy density operator behaves as dark energy and other terms can act as additional matter contents.}, which can depend on $\hat{H}^2(t,\vec{x})$.

A natural hypothesis for the form of the space dependence of dark-energy correlators is a power law behavior, which is in fact realised in the simple model that we discuss below.

In order to establish whether a dark energy of quantum origin could be enough to relieve the Hubble tension, we evaluate the conditional probability for local measurements (averaged over a spatial volume $V_1$) of the squared Hubble parameter to find a value above a threshold $H_1^2$, given that global measurements (averaged over a larger volume $V_2$ containing $V_1$) find a value below $H_2^2$. The square of the Hubble parameter is 
a natural fundamental variable to consider, as it appears in the Friedmann equation~(\ref{eq:pointH2general}).
We write the aforementioned conditional probability as
\begin{equation}
\label{eq:prob}
%\boxed{
{\rm P}\left(\left[H^2\right]_{V_1}>H_1^2\bigg|\left[H^2\right]_{V_2}<H_2^2\right)
%}
,
\end{equation}
where $\left[H^2\right]_{V_1}\equiv(1/V_1)\int_{V_1} d^3x \, H^2(t_0,\vec{x})$ and similarly for $V_2$.~\footnote{More precisely, we are implicitly defining a classical stochastic variable $H^2(t_0,\vec{x})$, whose statistical properties are the same as those of the quantum operator $\hat{H}^2(t_0,\vec{x})$.} The notation $t_0$ indicates cosmological time today.
\\

\noindent The values of $H_1$ and $H_2$ are chosen as the mean values of the results from SH0ES and {\it Planck} mentioned before:
%\begin{eqnarray}
%\label{Riess_vs_Planck}
$
H_1=74.03~{\rm km/s~Mpc^{-1}},
$
%\nonumber\\
%\,,\quad 
$
H_2=67.4~{\rm km/s~Mpc^{-1}}.
$
%\end{eqnarray}
For the volumes $V_1$ and $V_2$, we assume spheres centered around the observer with radii $R_1$ and $R_2$ that we specify below. %in Section~\ref{sect:prob}.

\section{The model} 
\label{sect:model}

Let us consider a light spectator field during inflation that, for the sake of simplicity, will be taken as a scalar $\Phi$. The key ingredient of the class of theories that we want to study is the assumption that some mechanism determines the effective mass $M^2(t)$ of the spectator field in a cosmological background in such a way that, by the end of inflation, correlators reach super-Planckian values and that these correlators are then translated into an effective dark energy at late times. A simple realization of such a mechanism is discussed in~\cite{Glavan:2017jye} (see also~\cite{Glavan:2015cut,Glavan:2014uga,Glavan:2013mra}) and consists in a non-minimally coupled massive scalar field with action
\begin{equation}
S[\Phi] \!=\!\! \int\! d^D\!x  \sqrt{-g}
	\biggl\{\! -\frac{1}{2} g^{\mu\nu} \partial_\mu \Phi \partial_\nu \Phi
		\!-\! \frac{1}{2}m^2\Phi^2 \!-\! \frac{1}{2}\xi R \Phi^2\! \biggr\}
,
\label{eq:action}
\end{equation}
where R is the Ricci curvature scalar of the metric $g_{\mu\nu}$ 
and $g={\rm det}[g_{\mu\nu}]$.
It is clear that here the effective mass of the scalar field, $M^2=m^2+\xi R$, receives contributions from the bare mass $m$ and from a non-minimal coupling to the metric with coefficient $\xi$. As predicted in~\cite{Glavan:2017jye}, in this model the field correlators of a light scalar grow during inflation and their enhancement is larger when the non-minimal coupling is negative and it dominates over the bare mass. Other mechanisms contributing to  an effective mass are conceivable, like dynamical mass generation {\it via} interaction with another field acquiring a non-zero vacuum expectation value in a symmetry-breaking scenario. In the general case the effective mass is a field operator with a non-trivial derivative structure.

Notice that assuming a quadratic action for the spectator field at the classical level ensures a minimal amount of non-Gaussianity. However, at the level of quantum effective action, interactions with the metric perturbations (or in general with other fields) introduce additional 
non-Gaussianities in the evolution of the spectator scalar field, 
{\it e.g.} by generating effective self-interactions.
\\

\noindent We consider a spatially flat FLRW Universe in $D=4$ spacetime dimensions with scale factor $a(t)$, {\it i.e.} 
${\rm d}s^2=-{\rm d}t^2+a^2(t){\rm d}\vec{x}^2$, 
where $t$ is cosmological time and $\vec{x}$ are comoving coordinates.

The spectator scalar field is $\Phi(x)=\Phi(t,\vec{x})$ and its action, computed by specializing~(\ref{eq:action}) to the FLRW metric, is of the form 
\begin{eqnarray}
S[\Phi]&\equiv&\int\! {\rm d}t{\rm d}^3x \, \mathcal{L}_\Phi=\int\! {\rm d}t{\rm d}^3x \, a^3
\label{scalar field action}\\
&&
\hskip -1.cm
\times
\left\{\frac12 \dot{\Phi}^2-\frac{1}{2a^2}\bigl(\vec{\nabla}\Phi\bigr)^2-\frac12 \left[m^2+6\xi (2-\epsilon) H^2\right]\Phi^2\right\} 
,
\nonumber
\end{eqnarray}
where the dot ($\dot{~}$) denotes the derivative with respect to cosmological time ${\rm d}/{\rm d}t$ and we introduced $\epsilon\equiv-\dot{H}/H^2$.
After introducing the canonical momentum $\Pi(x) =a^3 \dot{\Phi}(x)$, quantization of the model goes in the standard way (see {\it e.g.}~\cite{Glavan:2017jye}) by promoting field and momentum to operators $\hat{\Phi}$ and $\hat{\Pi}$ obeying canonical commutation relations.

\noindent The effect of the scalar field on cosmological evolution is determined by its energy-momentum tensor 
$T_{\mu\nu}=(-2/\sqrt{-g})\delta S[\Phi]/\delta g^{\mu\nu}$, 
which in its operator form reads,  
\begin{eqnarray}
\hat{T}_{\mu\nu} 
	& = & \partial_\mu \hat{\Phi} \, \partial_\nu\hat{\Phi}
		- \frac{1}{2} g_{\mu\nu} g^{\alpha\beta} 
			\partial_\alpha \hat{\Phi} \, \partial_\beta \hat{\Phi}
		- \frac{m^2}{2} g_{\mu\nu} \hat{\Phi}^2
\nonumber\\
&+& \xi \bigl[ G_{\mu\nu}  - \nabla_\mu \nabla_\nu 
			+ g_{\mu\nu} \square \bigr]
			\hat{\Phi}^2 
\, ,
\label{energy_momentum}
\end{eqnarray}
where $G_{\mu\nu}$ is the Einstein curvature tensor.

Friedmann equations tell us that quantum fluctuations of the spectator field enter the expansion history of the Universe through expectation values of energy density $\rho_Q  \equiv \langle\hat{\rho}_Q\rangle\equiv  - \langle \hat{T}^0{}_0\rangle$ and pressure $p_Q \delta^i_j  \equiv  \langle \hat{T}^i{}_j \rangle$, evaluated on a homogeneous and isotropic state that we take to be the vacuum $|\Omega\rangle$. Neglecting spatial gradients of the field, the energy density operator from~(\ref{energy_momentum}) simplifies to
\begin{eqnarray}
\label{local_rho_Q}
\hat{\rho}_Q(t,\vec{x})\!\!&\equiv&\!\!-\hat{T}^0{}_0(t,\vec{x})
=\frac{H^2}{2}\biggl\{\Bigl[ \Bigl( \frac{m}{H} \Bigr)^{\!2}
\!\!+\!6\xi\Bigr] \hat{\Phi}^2 (t,\vec{x})
\nonumber\\
\!\!&&\!\!\hskip -1.cm
+\,\frac{6\xi}{a^3 H}\left\{\hat{\Phi} (t,\vec{x}),\hat{\Pi} (t,\vec{x})\right\}\!+\!\frac{1}{a^6 H^2}\hat{\Pi}^2 (t,\vec{x})\biggr\}
.\qquad
\end{eqnarray}
In ~\cite{Glavan:2017jye}, the relevant expectation values for Friedmann equations are coincident correlators,
 $\Bigl\langle \hat{\Phi}^2 (t,\vec{x}) \Bigr\rangle$, $\Bigl\langle \hat{\Pi}^2 (t,\vec{x}) \Bigr\rangle$ and $\Bigl\langle \bigl\{ \hat{\Phi}(t,\vec{x}) ,\hat{\Pi}(t,\vec{x}) \bigr\} \Bigr\rangle$, where $\{\hat{A},\hat{B}\}\equiv\hat{A}\hat{B}+\hat{B}\hat{A}$. The result is that quantum fluctuations of the spectator field behave at late times as an effective {\it dark energy}.

Here we want to investigate the consequences of the spatial correlations inherited by dark energy. Therefore we turn our attention to 4-point functions built from field and momentum operators which are relevant for density-density correlators.
%\footnote{Spatial gradients of the field operator are also present in the energy-momentum tensor operator~(\ref{energy_momentum}). 
%Their effect on the evolution of 4-point functions can be neglected if $16|\xi|\ll1$, which is what we assume in the present analysis.}

\subsection{Stochastic formalism}
\label{Stochastic formalism}

\noindent The stochastic formalism of Alexey Starobinsky makes use of 
the fact that on super-Hubble scales non-conformal quantum fields
($\hat\Phi$, $\hat\Pi$)
grow during inflation, such that their correlators can be accurately described by their classical stochastic equivalents ($\hat\phi$, $\hat\pi$). 
These infrared fields obey, 
\begin{eqnarray}
\frac{{\rm d}}{{\rm d}t} \hat{\phi}(t,\vec{x}) 
	- a^{-3}\, \hat{\pi}(t,\vec{x})
	\!\!&=&\!\! \hat{f}_\phi(t,\vec{x}) \, ,
\label{EOMdPhiLong}
\\
a^{-3}\, \frac{{\rm d}}{{\rm d}t} \hat{\pi}(t,\vec{x})
	%- \frac{\nabla^2}{a^2} \hat{\phi}(t,\vec{x})
	+ M^2(t)\, \hat{\phi}(t,\vec{x})
	\!\!&=&\!\!  a^{-3} \hat{f}_\pi (t,\vec{x}) \, ,
\qquad
\label{EOMdPiLong}
\end{eqnarray}
where $\hat{f}_\phi(t,\vec{x})$ and $\hat{f}_\pi(t,\vec{x})$ are stochastic
sources, which arise because of coupling between
the infrared and ultraviolet modes (see {\it e.g.}~\cite{Glavan:2017jye}
for the precise definition of stochastic sources).

\subsection{IR correlators}
\label{subsect:corr}

We can now proceed to extend the treatment of~\cite{Glavan:2017jye} to study spatial correlations of long modes. Let us recall that, in eqs.~(30--32) of that paper, three 2-point functions were introduced because they appear in the expectation value of the energy-momentum tensor (conveniently rescaled with a power of $a^3(t) H(t)$ for each occurrence of a momentum operator), namely
$\Delta_{\phi\phi}(t) \equiv \Bigl\langle \hat{\phi}^2 (t,\vec{x}) \Bigr\rangle$, then $\Delta_{\phi\pi}(t) \equiv \Bigl\langle \bigl\{ \hat{\phi}(t,\vec{x}) ,\hat{\pi}(t,\vec{x}) \bigr\} \Bigr\rangle/[a^3(t)H(t)]$ and $\Delta_{\pi\pi}(t)\equiv\Bigl\langle \hat{\pi}^2 (t,\vec{x}) 
	\Bigr\rangle/[a^6(t)H^2(t)]$.
%
%\begin{eqnarray}
%\Delta_{\phi\phi}(t) & \equiv & 
%	\Bigl\langle \hat{\phi}^2 (t,\vec{x}) \Bigr\rangle \, ,
%\label{DeltaPhiPhi}
%\\
%\Delta_{\phi\pi}(t) & \equiv & 
%	\frac{1}{a^3(t) H(t)}\Bigl\langle \bigl\{ \hat{\phi}(t,\vec{x}) ,
%						\hat{\pi}(t,\vec{x}) \bigr\} \Bigr\rangle \, ,
%\nonumber
%\label{DeltaPhiPi}
%\\
%\Delta_{\pi\pi}(t) & \equiv & 
%	\frac{1}{a^6(t) H^2(t)} \Bigl\langle \hat{\pi}^2 (t,\vec{x}) 
%	\Bigr\rangle \, .
%\label{DeltaPiPi}
%\end{eqnarray}
%
Here we are interested in spatial fluctuations, therefore the relevant quantity is the density-density correlator at two different comoving coordinates $\vec{x}_1$ and $\vec{x}_2$ at the same cosmological time $t$:
$
%\begin{equation}
\langle \hat{\rho}_Q(t,\vec{x}_1) \hat{\rho}_Q(t,\vec{x}_2)\rangle
,
$
%\end{equation}
%
where the energy density operator is $\hat{\rho}_Q=-\hat{T}^0{}_0$. Since each $\hat{T}^0{}_0$ is quadratic in field/momentum operators, we infer that $\langle \hat{\rho}(t,\vec{x}_1) \hat{\rho}(t,\vec{x}_2)\rangle$ depends on 4-point functions of fields/momenta.
Due to the homogeneity and isotropy of the FLRW background, comoving coordinates $\vec{x}_1$ and $\vec{x}_2$ only appear in correlators through their relative comoving distance $r\equiv\|\vec{x}_2-\vec{x}_1\|$.

The relevant correlators are six symmetric combinations~(see \cite{inprep:2021etc}), $\Delta_{\phi^2,\phi^2}$, $\Delta_{\phi^2,\phi\pi}$, $\Delta_{\phi\pi,\phi\pi}$, $\Delta_{\phi^2,\pi^2}$, $\Delta_{\phi\pi,\pi^2}$, $\Delta_{\pi^2,\pi^2}$. The simplest one is
\begin{equation}
\Delta_{\phi^2,\phi^2}(t,r)  \equiv 
	\Bigl\langle \hat{\phi}^2 (t,\vec{x}_1) \hat{\phi}^2 (t,\vec{x}_2) \Bigr\rangle \,.
\label{DeltaPhi2,Phi2}
%\\
%\Delta_{\phi^2,\phi\pi}(t,r) & \equiv & 
%	\frac{1}{a^{3}(t) H(t)}\Bigl\langle \hat{\phi}^2 (t,\vec{x}_1)\left\{\hat{\phi} (t,\vec{x}_2),\hat{\pi} (t,\vec{x}_2)\right\}
%\nonumber\\	
%&+&\left\{\hat{\phi} (t,\vec{x}_1),\hat{\pi} (t,\vec{x}_1)\right\}\hat{\phi}^2 (t,\vec{x}_2) \Bigr\rangle \, ,
%\nonumber
%\label{DeltaPhi2,PhiPi}
%\\
%\Delta_{\phi\pi,\phi\pi}(t,r) \!\!& \equiv &\!\! 
%	\frac{\Bigl\langle\! \left\{\hat{\phi} (t,\vec{x}_1),\hat{\pi} (t,\vec{x}_1)\right\} \left\{\hat{\phi} (t,\vec{x}_2),\hat{\pi} (t,\vec{x}_2)\right\}\! \Bigr\rangle}{a^{6}(t) H^2(t)}
% ,
%\nonumber
%\label{DeltaPhiPi,PhiPi}
%\\
%\Delta_{\phi^2,\pi^2}(t,r) \!\!& \equiv &\!\!  
%	\frac{\!\Bigl\langle\!\hat{\phi}^2 (t,\vec{x}_1) \hat{\pi}^2 (t,\vec{x}_2)
%	  \!+\!\hat{\pi}^2 (t,\vec{x}_1) \hat{\phi}^2 (t,\vec{x}_2)\!\Bigr\rangle}{a^{6}(t) H^2(t)} 
%,
%\nonumber
%\label{DeltaPhi2,Pi2}
%\\
%\Delta_{\phi\pi,\pi^2}(t,r) & \equiv & 
%	\frac{1}{a^{9}(t) H^3(t)}\Bigl\langle \hat{\pi}^2 (t,\vec{x}_1)\left\{\hat{\phi} (t,\vec{x}_2),\hat{\pi} (t,\vec{x}_2)\right\}
%5\nonumber\\	
%&+&
%\left\{\hat{\phi} (t,\vec{x}_1),\hat{\pi} (t,\vec{x}_1)\right\}\hat{\pi}^2 (t,\vec{x}_2) \Bigr\rangle \, ,
%\nonumber
%\label{DeltaPhiPi,Pi2}
%\\
%\Delta_{\pi^2,\pi^2}(t,r) & \equiv & 
%	\frac{1}{a^{12}(t) H^4(t)} \Bigl\langle \hat{\pi}^2 (t,\vec{x}_1) \hat{\pi}^2 (t,\vec{x}_2) 
%	\Bigr\rangle \, .
%\nonumber
%\label{DeltaPi2,Pi2}
\end{equation}
We can study the time evolution of 4-point functions and identify their stochastic noise sources induced by the stochastic forces $\hat{f}_\phi(t,\vec{x})$ and $\hat{f}_\pi(t,\vec{x})$  of~(\ref{EOMdPhiLong}--\ref{EOMdPiLong}), to obtain a system of six coupled differential equations. Here we give two equations, 
\begin{eqnarray}
\frac{{\rm d}}{{\rm d}N} \Delta_{\phi^2,\phi^2} \!-\! \Delta_{\phi^2,\phi\pi} \!=\! n_{\phi^2,\phi^2} 
,\quad
\label{eomPhi2,Phi2}
\\
\frac{{\rm d}}{{\rm d}N} \Delta_{\pi^2,\pi^2} \!+\! 4(3\!-\!\epsilon) \Delta_{\pi^2,\pi^2}
	\!+\! \Bigl( \frac{M}{H} \Bigr)^{\!2} 
		\Delta_{\phi\pi,\pi^2} \!=\! n_{\pi^2,\pi^2} 
,\quad
\label{eomPi2,Pi2}
\end{eqnarray}
where we have traded the time $t$ for the number of e-foldings 
$N\equiv\ln a(t)$, and the source in Eq.~(\ref{eomPhi2,Phi2}) is,
\begin{eqnarray}
n_{\phi^2,\phi^2}(t,r)  &=&  \frac{1}{H(t)} \Bigl\langle 
	\left\{ \hat{f}_{\phi}(t,\vec{x}_1) ,\hat{\phi}(t,\vec{x}_1) \right\} \hat{\phi}^2(t,\vec{x}_2)
\nonumber\\
&+&
          \hat{\phi}^2(t,\vec{x}_1) \left\{ \hat{f}_{\phi}(t,\vec{x}_2) ,\hat{\phi}(t,\vec{x}_2) \right\} 
		\Bigr\rangle 
\,
\qquad (r=\|\vec x_1-\vec x_2\|)
 .\quad
\label{n phi2,phi2}
\end{eqnarray}
The complete set of equations and sources is found 
in~\cite{inprep:2021etc}.
The stochastic sources can be computed in terms of the mode function
$\varphi(t,k)$ of $\hat \Phi$. The source in~(\ref{n phi2,phi2}) is thus,
\begin{eqnarray}
\label{eq:noise}
 n_{\phi^2,\phi^2} &=& \frac{1}{2\pi^4} 
(\mu aH)^3(1-\epsilon)\left[|\varphi(t,k)|^2\right]_{k=\mu aH}
%\\&\times &\!\!
\int_{k_0}^{\mu aH}{\rm d}k~k^2 |\varphi(t,k)|^2\left[1+2j_0(\mu aHr)j_0(kr)\right] 
,\quad
%\nonumber
\end{eqnarray}
where $j_0(kr)\equiv\sin(kr)/(kr)$ is the spherical Bessel function
and $\mu aH$ ($\mu\ll 1$) is an UV cut-off. 
% of order zero evaluated at $kr$.

\subsection{Time evolution of correlators}
\label{timecorrelators}

Analogously as it was done in~\cite{Glavan:2017jye},
one can solve the equations of motion for the correlators 
(\ref{eomPhi2,Phi2}--\ref{eomPi2,Pi2}) from an early inflation,
through radiation and matter era. Even though the correct evolution
of the correlators in matter era requires taking account of 
the dark energy backreaction,  for simplicty here we model
the late time universe by non-relativistic matter and cosmological constant.
This introduces a few percent error~\cite{Glavan:2017jye}, but has 
the advantage that one can solve the evolution of the relevant
four-point functions exactly. The result is~\cite{inprep:2021etc}
\begin{eqnarray}
\Delta_{(\rm 4)}(N_0,r) \!\!&\equiv&\!\!
 \left(\!\Delta_{\phi^2,\phi^2},\!\Delta_{\phi^2,\phi\pi},\!\Delta_{\phi\pi,\phi\pi},\!\Delta_{\phi^2,\pi^2},\!\Delta_{\phi\pi,\pi^2},\!\Delta_{\pi^2,\pi^2}\!\right)
 \nonumber\\
                      &&\hskip -2cm
\simeq\frac{H_I^4e^{16|\xi|N_I}e^{8\zeta N_0}}{1024\pi^4\xi^2}s(r)
(1, 8\zeta, 16\zeta^2, 8\zeta^2, 32\zeta^3, 16\zeta^4)
,
\label{4pt_final}
\end{eqnarray}
where $N_0=\ln{\left(\frac{\Omega_M}{\Omega_R}\right)} \simeq8.1$ 
e-foldings since matter-radiation equality,
\begin{equation}
\zeta\equiv|\xi|-\frac{\frac{1}{6\Omega_\Lambda}
\!\left(\frac{m}{H_0}\right)^2-|\xi|}{\frac{3}{2}\frac{\ln(\Omega_R)}{\ln(\Omega_M)}-1}
\,,
\end{equation}
and $s(r)$ contains spatial dependence, which is well approximated by,
\begin{equation}
\label{spatial_inflation}
s(r)\simeq 
\begin{cases}
      3 & \text{if $0\leq \mu a_{\rm in}H_I r< {\rm e}^{-N_I}$} \\
      3-2\left(\mu a_{\rm in}H_I r\right)^{16|\xi|} & \text{if ${\rm e}^{-N_I}< \mu a_{\rm in}H_I r< 1$} \\
      1 & \text{if $\mu a_{\rm in}H_I r\geq 1$}\,,
\end{cases} 
\end{equation}
where $a_{\rm in}$ is the scale factor at the beginning of inflation. Therefore the comoving length scale ruling spatial variations is the quantity $(\mu a_{\rm in}H_I)^{-1}$, {\it i.e.} the comoving Hubble horizon at the beginning of inflation (up to the $\mu^{-1}$ factor).
Furthermore, $\Omega_R$, $\Omega_M$ and $\Omega_\Lambda\simeq 1-\Omega_M$  are the energy density fractions in radiation and 
matter today, taken to be 
$\Omega_R=9.1\cdot10^{-5}$ and $\Omega_M=0.3$. 

\noindent 
The quantum backreaction of the scalar field contributes to energy density and pressure during matter domination as in~\cite{Glavan:2017jye}, 
\begin{eqnarray}
\rho_Q &=& \frac{H_I^2}{32\pi^2|\xi|}e^{8|\xi|N_I} 
	\biggl(\frac{m^2}{2}-3|\xi|H^2\biggr)\, , 
\qquad\nonumber\\
p_Q &=& - \frac{H_I^2}{32\pi^2|\xi|}e^{8|\xi|N_I}\frac{m^2}{2} 
\, .
\label{mat rho p}
\end{eqnarray}
Matching the cosmological constant-like term to its value today 
$\Omega_\Lambda$ determines the number of inflationary e-foldings 
$N_I$ in terms of the model parameters $m$ and $|\xi|$,
\begin{equation}
N_I = \frac{1}{8|\xi|}  \ln \biggl[ 
	24\pi|\xi| \Bigl( \frac{m_P}{H_I} \Bigr)^2
	\Bigl( \frac{H_{DE}}{m} \Bigr)^2 
	\biggr] \, ,
\label{inflation_efolds_relation}
\end{equation}
where $m_P\equiv G^{-1/2}$ is the Planck mass, $H_{\rm DE}\equiv H_0\sqrt{\Omega_\Lambda}$ 
and $H_0$ is the Hubble parameter today. 

Dark energy eventually leads the cosmological expansion provided that the cosmological constant-like term dominates over the matter-like term in the quantum backreaction until late times, {\it i.e.} we require
\begin{equation}
|\xi|<\frac{1}{6}\left(\frac{m}{H_{\rm DE}}\right)^2
\qquad {\rm and} \quad \xi<0
\,.
\label{limits_xi}
\end{equation}
We also need $m/{H_{\rm DE}}<1$ so that the scalar field stays light throughout the cosmological expansion.
\\

\noindent The spatial dependence in~(\ref{4pt_final})
 is the one inherited from inflation through the function $s(r)$
 in~(\ref{spatial_inflation}). Before discussing the Hubble tension problem, it is useful to work on the comoving length scale 
$(\mu a_{\rm in}H_I)^{-1}$ associated to $s(r)$  in order to relate its physical size today to the  model parameters. Assuming an instantaneous reheating after inflation, we find the following ratio between the physical length scale of spatial variations today and the current Hubble horizon:
\begin{equation}
\label{spatial_ratio}
\frac{(\mu a_{\rm in}H_I)^{-1}a_0}{H_0^{-1}}=\mu^{-1}e^{N_I} \left(\frac{H_I}{H_0}\right)^{-\frac12}\Omega_R^{-\frac14}
\,.
\end{equation}
Spatial variations are relevant for the Hubble tension problem discussed in section~(\ref{sect:prob}) 
only if they are significant at sub-Hubble scales. Therefore, we require that the ratio appearing in the left-hand side of~(\ref{spatial_ratio}) is at most of order one, which then limits the number of inflationary e-foldings. For $H_I\simeq 10^{13}~{\rm GeV}$, $H_0\simeq 10^{-33}~{\rm eV}$ (and $\mu$ not too small), this gives
%\begin{equation}
%\label{limit_60_efolds}
$
N_I\lesssim 60.
$
%\end{equation}
%
The condition on $N_I$ translates into a requirement on the parameters of the model, because $N_I$ is determined by $|\xi|$ and $m$ via the combination in~(\ref{inflation_efolds_relation}).

\section{Relieving the Hubble tension}
\label{sect:prob}

Note that the local expansion rate operator $\hat{H}^2(t,\vec{x})$ in~(\ref{eq:pointH2general}) can be recast as,
\begin{equation}
\hat{H}^2(t_0,\vec{x})\!=\!\frac{1}{3M_P^2}
\frac{\rho_{C}(t_0)\unit \!+\!\frac{m^2}{2}\hat{\phi}^2(t_0,\vec{x})}
       {\unit\!+\!\frac{|\xi|}{M_P^2}\hat{\phi}^2(t_0,\vec{x})}
%\left[\rho_{C}(t_0)\unit
%\!+\!\frac{m^2}{2}\hat{\phi}^2(t_0,\vec{x})\right]
%\left[\unit\!+\!\frac{|\xi|}{M_P^2}\hat{\phi}^2(t_0,\vec{x})\right]^{-1}
\,,
\label{eq:pointH2today}
\end{equation}
where we made use of $\hat \rho_Q$ from~(\ref{local_rho_Q}) and neglected contributions from of $\{\hat{\phi},\hat{\pi}\}$ and $\hat{\pi}^2$, which is a reasonable assumption~\cite{inprep:2021etc}.
Next, we assume that $H_0\simeq H_2=67.4~{\rm km/s~Mpc^{-1}}$, meaning that the statistical average value $H_0$ is estimated very well by the result of global measurements.

\noindent The translation of a probability statement for $\hat{H}^2$ like that in~(\ref{eq:prob}) into an equivalent statement for $\hat{\phi}^2$ depends on whether these two quantities are related by a growing or decreasing function.
A simple analysis of~(\ref{eq:pointH2today}) shows that  $\frac{d(\hat{H}^2)}{d(\hat{\phi}^2)}
\propto \frac{m^2}{2}-\frac{|\xi|}{M_P^2}\rho_{C}(t_0)\propto \left[1-6|\xi|\left(\frac{H_0}{m}\right)^2\right]$,
such that $\hat{H}^2$ is a growing (decreasing) function of $\hat{\phi}^2$ when 
$|\xi|<\frac{1}{6}\left(m/H_0\right)^2$  ($|\xi|>\frac{1}{6}\left(m/H_0\right)^2$).
 For example, with the choice $\xi=-\frac{1}{6}\left({m}/{H_{\rm DE}}\right)^2$ that saturates the constraint~(\ref{limits_xi}), the quantity $1-6|\xi|\left({H_0}/{m}\right)^2=1-\Omega_\Lambda^{-1}<0$ and 
$\hat{H}^2$ is a decreasing function of $\hat{\phi}^2$.
\\

\noindent 
In what follows, we introduce some simplifications, whose 
accuracy is discussed in detail in Ref.~\cite{inprep:2021etc}.
Firstly, we shall assume that the field $\hat{\psi}(t_0,\vec{x})\equiv\hat{\phi}^2(t_0,\vec{x})$ is a classical stochastic variable 
${\psi}(t_0,\vec{x})$
which is Gaussian distributed. Secondly, we shall assume that 
the stochastic properties of $\psi$ can be modeled by 
a Gaussian probability distribution, 
\begin{eqnarray}
P[\psi]&\propto&\exp\bigg\{-\frac12\int {\rm d}^3x\int {\rm d}^3y~\left[\psi(t_0,\vec{x})-\psi_0\right]
\nonumber\\
&&\times\, C^{-1}(t_0,\|\vec{x}-\vec{y}\|)\left[\psi(t_0,\vec{y})-\psi_0\right]\bigg\}
\,,
\label{prob_dist}
\end{eqnarray}
where the position-independent mean value is $\psi_0=\Delta_{\phi\phi}(t_0)$, while the covariances are
\begin{equation}
C(t_0,\|\vec{x}-\vec{y}\|)=\Delta_{\phi^2,\phi^2}(t_0,\|\vec{x}-\vec{y}\|)-\Delta_{\phi\phi}^2(t_0)\,. 
\end{equation}
As for the domain of definition, the probability distribution $P[\psi]$ is assumed to be normalized in the global volume $V_2$ (which includes $V_1$).

Using the usual relation between conditional probabilities and joint probabilities, the problem of finding~(\ref{eq:prob}) amounts to study the joint probability distribution of the mean variables $[\psi]_{V_1}\equiv(1/V_1)\int_{V_1} d^3x \,\psi(t_0,\vec{x})$ and $[\psi]_{V_2}\equiv(1/V_2)\int_{V_2} d^3x \,\psi(t_0,\vec{x})$. These are still Gaussian variables and one can evaluate their ($2\times2$ and symmetric) covariance matrix
\begin{eqnarray}
{\mathcal M}\!=\!
    \begin{bmatrix}
    {\mathcal M}_{11} & {\mathcal M}_{12} \\
    {\mathcal M}_{12} & {\mathcal M}_{22} \\
    \end{bmatrix}
 \!,\;\;\;    {\mathcal M}_{ij} \!=\!\frac{1}{V_iV_j}\!\!\int_{V_i}\!\!{\rm d}^3x\!\! \int_{V_j}\!\!{\rm  d}^3y
~C(t_0,\|\vec{x}\!-\!\vec{y}\|)
 \nonumber
%\!\!&=&\!\!
% \begin{bmatrix}
 %   \frac{1}{V_1^2}\int_{V_1} d^3x \int_{V_1} d^3y\, C(t_0,\|\vec{x}-\vec{y}\|)& \frac{1}{V_1 V_2}\int_{V_1} d^3x \int_{V_2} d^3y\, C(t_0,\|\vec{x}-\vec{y}\|) \\
%   \frac{1}{V_1 V_2}\int_{V_1} d^3x \int_{V_2} d^3y\, C(t_0,\|\vec{x}-\vec{y}\|) & \frac{1}{V_2^2}\int_{V_2} d^3x \int_{V_2} d^3y\, C(t_0,\|\vec{x}-\vec{y}\|) \\
 %   \end{bmatrix}
\,.
\end{eqnarray}
An important quantity is the correlation coefficient $\rho$ defined as
%\begin{equation}
$
%\rho=\frac{{\mathcal M}_{12}}{\sqrt{{\mathcal M}_{11}{\mathcal M}_{22}}}\,.
\rho={\mathcal M}_{12}/\sqrt{{\mathcal M}_{11}{\mathcal M}_{22}}.
$
%\end{equation}
We can evaluate the covariance matrix $\mathcal M$ and the correlation coeffcient $\rho$ by using for the volumes $V_1$ and $V_2$ two spheres of radii $R_1$ and $R_2$ centered around the observer.  
The conditional probability~(\ref{eq:prob}) can be reduced to, 
\begin{eqnarray}
\label{eq:probfinal_zerodeltachi2}
&&\hskip -0.5cm
{\rm P}\left(\left[H^2\right]_{V_1}>H_1^2\bigg|\left[H^2\right]_{V_2}<H_2^2\right)
\\
&&\hskip 1.cm
=\,2~T\left(\sqrt{2}\delta\chi,\frac{-\rho}{\sqrt{1\!-\!\rho^2}}\right)
                  \!+\!\frac12\left[1\mp{\rm erf}\left(\delta\chi\right)\right]
,
\nonumber
\qquad
\end{eqnarray}
where 
$T(h,a)=\frac{1}{2\pi}\int_0^a {\rm d}x
\exp\Big(\!\!-\!\frac{h^2}{2}(1\!+\!x^2)\Big)/(1\!+\!x^2)$ 
is Owen's T function and 
\begin{equation}
\delta \chi \!= \!\frac{\frac{H_{1}^2}{H_0^2}\!-\!1}{\sqrt{2J_{11}}}
\frac{(m/H_{\rm DE})^2\!+\!6|\xi|}{(m/H_{0})^2
                      \!-\! 6|\xi|\frac{H_{1}^2}{H_0^2}}
,\quad J_{11} \!=\! I_{11} e^{8\zeta N_0}\!-\!1
,
\label{delta chi12}
\end{equation}
where $I_{11} = \int_0^2 {\rm d}x\Big(\frac{3}{16}x^5-\frac94x^3+3x^2\Big)s_1(x)$ and 
\begin{equation}
s_{1}(x)=  
\begin{cases}
      3 & \text{if $0\leq x\leq k_{1}~e^{-N_I}$} \\
      3-2\left(\frac{x}{k_{1}}\right)^{16|\xi|} & \text{if $k_{1}~e^{-N_I}< x\leq k_{1}$} \\
      1 & \text{if $x\geq k_{1}$}
\end{cases}
\,,
\label{s1}
\end{equation}
with $k_{1}={\rm e}^{N_I}\left(\frac{H_I}{H_0}\right)^{-\frac12}\Omega_R^{-\frac14}\left(\mu H_0 R_{1}\right)^{-1}$.

\begin{figure}[h]
\centering
\includegraphics[width=0.99\columnwidth]{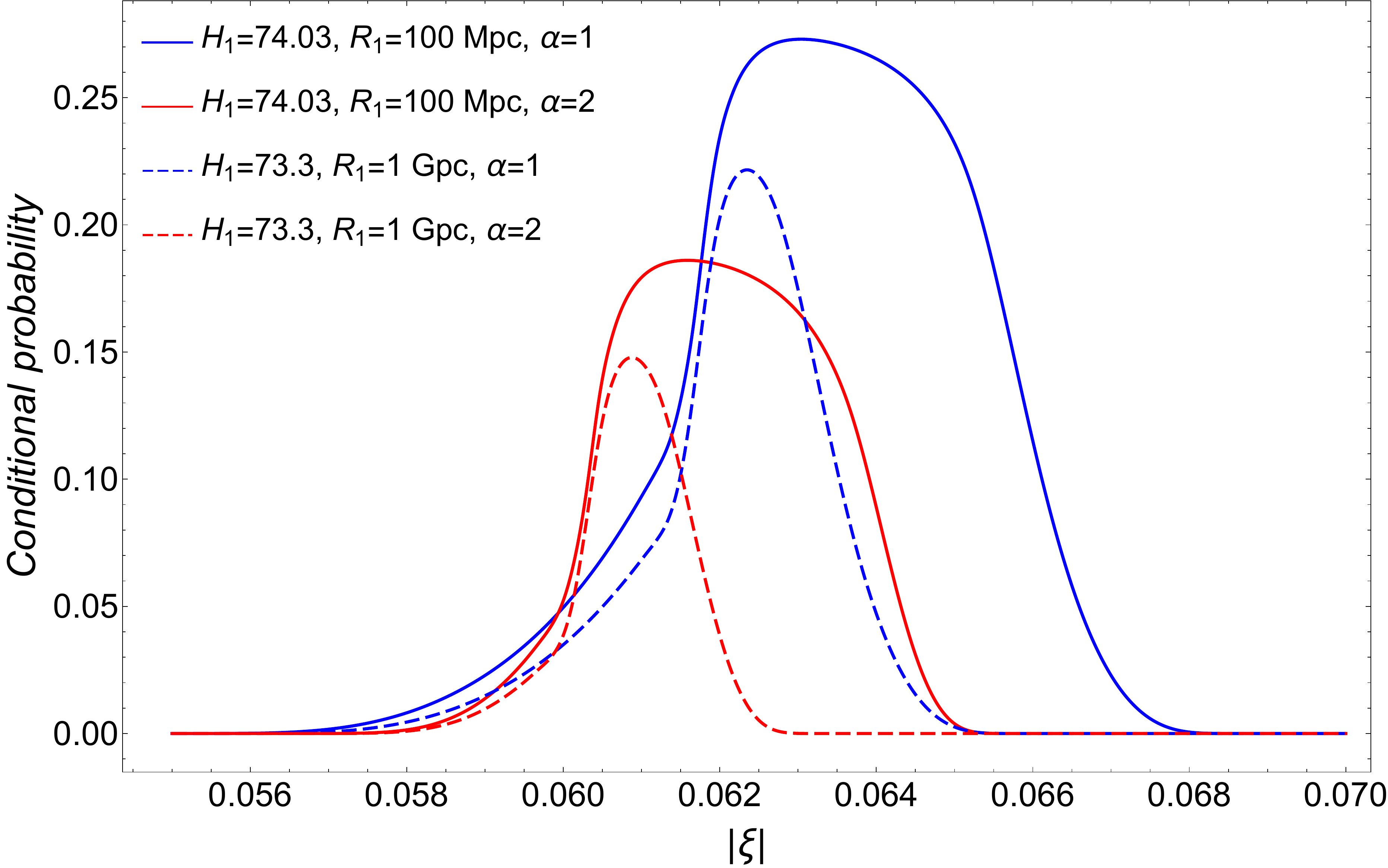}
\caption{The conditional probability in~(\ref{eq:probfinal_zerodeltachi2}), as a function of the non-minimal coupling $|\xi|$, for constant values of $\alpha=\frac{1}{6|\xi|}\left(m/H_{\rm DE}\right)^2$. The blue curves correspond to the limit case $\alpha=1$, which is its minimum possible value. The parameter $\xi$ is mildly tuned to a rather natural value, since the width of the region of $\xi$ values relevant for relieving the Hubble tension (to less than $2\sigma$) can reach $\Delta\xi/|\xi|\sim{\mathcal O}(10\%)$ in the best case.
%Note that the peaks in probability are rather broad, meaning that no fine tuning 
%in the parameters $|\xi|$ and $m/H_{\rm DE}$ is needed. 
\label{fig:prob}
}
\end{figure}

For definiteness we choose the radius $R_2$ of the global volume such that $\mu R_2= H_0^{-1}\simeq$ 4400 Mpc and the radius $R_1$ of the local volume such that $\mu R_1= $ 100 Mpc (we set $\mu=1$).
The conditional probability~(\ref{eq:probfinal_zerodeltachi2}) depends on the two parameters of the model $\xi$ and $m$. Solid lines in Figure~\ref{fig:prob} show its behavior as a function of $|\xi|$ for constant values of the ratio $\alpha\equiv\frac{1}{6|\xi|}\left({m}/{H_{\rm DE}}\right)^2$. Notice that the condition~(\ref{limits_xi}) imposes to choose $\alpha>1$. Furthermore the condition of light field $m/H_{\rm DE}<1$ means that, for a given constant $\alpha$, only values of $|\xi|$ with $|\xi|<1/(6\alpha)$ have to be considered (this is automatically true for the values of $\alpha$ and the plot range chosen in Figure~\ref{fig:prob}).
The number of inflationary e-foldings throughout the range of parameters in Figure~\ref{fig:prob} is of order $N_I\sim 60$ and is therefore compatible with the approximate constraint $N_I\lesssim 60$.
%in~\ref{limit_60_efolds}.

\begin{figure}[h]
\centering
\includegraphics[width=0.99\columnwidth]{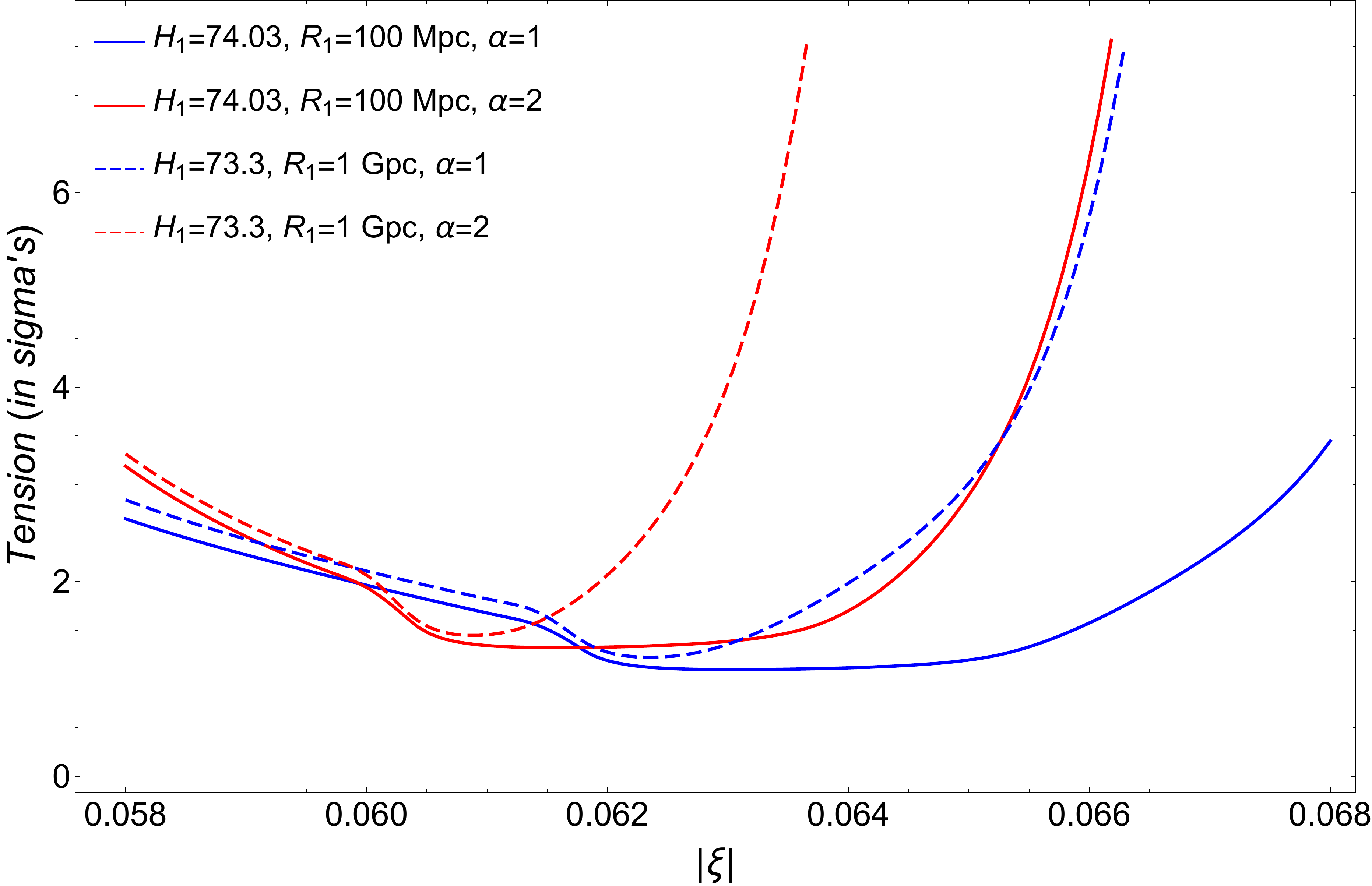}
\caption{Hubble tension in standard deviations $\sigma$ as a function of the non-minimal coupling $|\xi|$, in the same conditions for $m/H_{\rm DE}$ as in Figure~\ref{fig:prob}. The plot range is slightly zoomed in with respect to that in Figure~\ref{fig:prob}. The tension can be reduced down 
to $\simeq 1\sigma$ (solid lines) and $\simeq 1.2 \sigma$ (dashed lines)
for local measurements in $R_1 = 100~{\rm Mpc}$ and 
$R_1 = 1~{\rm Gpc}$, respectively.
\label{fig:tension}
}
\end{figure}

\noindent Figure~\ref{fig:tension} shows the corresponding Hubble tension zooming slightly on a more central region. As can be seen from the plot, the model is able to reduce the tension down to almost $1\sigma$ which is a great improvement with respect to the $>4\sigma$ tension in $\Lambda$CDM. This has been possible thanks to spatial fluctuations of dark energy inherited from inflation.

\noindent  In the limit of zero local volume one gets the largest conditional probability that can be reached, which can be obtained from~(\ref{eq:probfinal_zerodeltachi2}) to be $\frac{1}{2}-\frac{1}{\pi}{\rm arctan}\left(\frac{1}{\sqrt{2}}\right)\simeq 30.4\%$, corresponding to 
a ``tension" of only $1.03\sigma$. This is in agreement with the solid lines in Figures~\ref{fig:prob}--\ref{fig:tension}, where the ratio 
$R_1/R_2\simeq  0.02$ is small but finite. To mimic H0LiCOW measurements~\cite{Wong:2019kwg} we increase $R_1\simeq 1~{\rm Gpc}$ and set $H_1=73.3~{\rm km/s~Mpc^{-1}}$. The corresponding results are shown by the dashed lines in Figures~\ref{fig:prob}--\ref{fig:tension}. Our model relieves the Hubble tension, 
but now the probability peaks at about 20\%, corresponding to about 1.2$\sigma$ tension.

Another major problem in $\Lambda$CDM is the $\sigma_8$ tension. A full quantitative discussion of our model with regard to this issue goes beyond the scope of the paper. However, it is interesting to have a qualitative understanding of it. In our model, a higher local value of the Hubble parameter (with respect to the global one) means a positive local fluctuation in dark energy. Assuming for simplicity that there is no local curvature effect so that the sum of the local matter density fraction and local dark energy density fraction is still $1$, this would correspond to a reduced value of the local matter density fraction ($\Omega_M$) with respect to its global value. Smaller $\Omega_M$ means less growth of structures (e.g. assuming that the growth rate $f$ obeys the widely used power law $f=\Omega_M^\gamma$ with $\gamma\approx0.55$), thus going in the direction of reducing the amplitude of local matter perturbations. This is in agreement with current data, see e.g. Fig. 2 in~\cite{Troster:2019ean}. 

More work is needed to reveal and test the deviations of the model proposed here from $\Lambda$CDM. Some plots on the evolution of the Hubble rate and the dark energy equation of state in recent cosmological times were presented in Figs. 2, 3, 4, 5 of~\cite{Glavan:2017jye} for a few choices of the parameters $\xi$ and $m$; the expected deviations are within reach of Euclid and Vera Rubin Observatory.
In \cite{Demianski:2019vmq} the authors test the same quantum dark energy model against cosmological data and compare its performance to $\Lambda$CDM, finding that the model is slightly favored, although not at a statistically significant level.
We are currently investigating the form of luminosity distance correlators for type Ia supernovae induced by spatially correlated dark energy, which provide a test for deviations from $\Lambda$CDM. This is just an example of physical observables that could be used to test the features predicted by the model. It is also worth mentioning that, within a different approach using cosmological perturbations, dark energy with very large-scale inhomogeneities has been considered in~\cite{Nan:2021prt} and their amplitude is constrained with CMB anisotropies data.

\vspace{5mm}

\noindent
{\bf Acknowledgments.} 
The authors are grateful to Tanja Hinderer and Elisa Chisari for useful comments.
This work is part of the Delta ITP consortium, a program of the Netherlands Organisation for Scientific Research (NWO) that is funded by the Dutch Ministry of Education, Culture and Science (OCW) - NWO projectnumber 24.001.027.

\appendix

\section{Four-point functions calculation}
Here we provide some details on the calculation leading to the result~(\ref{4pt_final}) by following the evolution of correlators in the three cosmological epochs.

\subsection{Inflation}

The mode function $\varphi(t,k)$ to be used in exact de Sitter inflation ($\epsilon=0$) is the Chernikov-Tagirov-Bunch-Davies (CTBD) mode function
\begin{equation}
\varphi(t,k) = \sqrt{\frac{\pi}{4a^3H_I}} \ 
	H_{\nu_I}^{(1)} \Bigl( \frac{k}{aH_I} \Bigr) \, ,
\qquad
\nu_I \equiv \sqrt{\tfrac{9}{4}-\bigl( \tfrac{M}{H_I} \bigr)^{\!2} } \, ,
\label{mode function}
\end{equation}
where $H_I$ is the constant inflationary Hubble rate, $M^2=m^2 +12\xi H_I^2$ is the constant squared effective mass and $H_{\nu_I}^{(1)}$ is the Hankel function of the first kind.
The CTBD mode function is needed to specify the sources in the  system of first-order differential equations including~(\ref{eomPhi2,Phi2}--\ref{eomPi2,Pi2}). Since we are dealing with long-wavelength modes in the integral of~(\ref{eq:noise}), it is sufficient to consider an IR approximation of the mode function
\begin{equation}
\varphi(t,k) \approx \frac{-i}{\sqrt{\pi}}2^{\nu_I-1} \Gamma(\nu_I)a^{\nu_I-3/2}H_I^{\nu_I-1/2}k^{-\nu_I}\,.
\label{mode function approx}
\end{equation}
As initial conditions, we assume correlators to be zero at the beginning of inflation.\footnote{The effect of non-zero initial conditions for 2-point functions (and therefore for 4-point functions as well) inherited from a pre-inflationary epoch will be discussed in~\cite{inprep:2021etc}.}
During de Sitter inflation, the evolution with the number of e-foldings $N$ of correlators $\Delta_{(\rm 4)}(N,r)$ defined by $\Delta_{(\rm 4)}(N,r) \!\!\equiv\!\!
 \left(\!\Delta_{\phi^2,\phi^2},\!\Delta_{\phi^2,\phi\pi},\!\Delta_{\phi\pi,\phi\pi},\!\Delta_{\phi^2,\pi^2},\!\Delta_{\phi\pi,\pi^2},\!\Delta_{\pi^2,\pi^2}\!\right)$ can be read from equations like ~(\ref{eomPhi2,Phi2}--\ref{eomPi2,Pi2}) with $\epsilon=0$, resulting in the system of equations
\begin{equation}
\frac{{\mathrm d}}{{\mathrm d} N}\Delta_{(\rm 4)}(N,r)+A~\Delta_{(\rm 4)}(N,r)=n(N,r)\,,
\label{system_deSitter}
\end{equation}
where $A$ is the following $6\times6$ constant matrix containing $X\equiv(M/H_I)^2=(m/H_I)^2+12\xi$,
\begin{equation}
A=
\begin{bmatrix}
   0 & -1 & 0 & 0 & 0 & 0 \\
 4X & 3 & -2 & -2 & 0 & 0 \\
 0 & 2X & 6 & 0 & -2 & 0 \\
 0 & X & 0 & 6 & -1 & 0 \\
 0 & 0 & 2X & 2X & 9 & -4 \\
 0 & 0 & 0 & 0 & X & 12 \\
\end{bmatrix}\,,
\end{equation}
and the source vector $n(N,r)$ can be evaluated from the mode function. For\footnote{As already shown in~\cite{Glavan:2017jye}, this hierarchy choice for the parameters gives the best enhancement of correlators in inflation, which also means that a smaller number of inflationary e-foldings is required to match the dark energy content of the Universe today.} $\xi<0$ and $(m/H_I)^2\ll|\xi|\ll 1$, working at the lowest order in the non-minimal coupling $\xi$, we get (recalling that the scale factor is $a(N)={\mathrm e}^N$)
\begin{equation}
n(N,r)=\left(\frac{H_I}{2\pi}\right)^4 \left[1+2j_0(\mu {\mathrm e}^N H_I r)\right]\left(\frac{1}{4|\xi|},4,16|\xi|,8|\xi|,64\xi^2,64|\xi|^3\right)\,.
\label{noise_deSitter}
\end{equation}

The differential problem~(\ref{system_deSitter}) is conveniently solved by diagonalizing the matrix $A$. Denoting by $\lambda_{j}$ the eigenvalues of $A$, by $B$ the change-of-basis matrix (matrix whose columns are the eigenvectors of $A$) and by $B^{-1}$ its inverse, the $i$-th component of the correlators vector $\Delta_{(\rm 4)}(N,r)$ is given by

\begin{equation}
\Delta_{({\rm 4}),i}(N,r)=\sum_{j}B_{ij}e^{-\lambda_j N}\int_0^N dN'~e^{\lambda_j N'}\sum_{k}(B^{-1})_{jk}~n_k(N',r)\,.
\label{sol_deSitter}
\end{equation}
The integral can be evaluated by approximating the spherical Bessel function $j_0(x)=\sin(x)/x$ appearing in~(\ref{noise_deSitter}) as $j_0(x)\simeq\theta(1-x)$ where $\theta$ is the Heaviside step function. Furthermore, in the limit of small $X$ it can be shown that the sum over eigenvalues in~(\ref{sol_deSitter}) is dominated by the smallest eigenvalue of $A$ which is $6-2\sqrt{9-4X}=4/3~X+{\mathcal O}(X^2)$. With these simplifications one finds that, at the end of inflation (lasting $N_I$ e-foldings), the 4-pt correlators are
\begin{equation}
\Delta_{(\rm 4)}(N_I,r)\simeq\frac{H_I^4}{1024\pi^4\xi^2}e^{16|\xi|N_I} s(r)\left(1,16|\xi|,64\xi^2,32\xi^2,256|\xi|^3,256\xi^4\right)\,,
\label{final_deSitter}
\end{equation}
where the function $s(r)$ encoding spatial dependence was defined in~(\ref{spatial_inflation}).

\subsection{Radiation epoch}
After the end of inflation, the correlators will evolve in radiation era starting from the initial conditions~(\ref{final_deSitter}) inherited from inflation. Similarly to the case of 2-pt functions studied in~\cite{Glavan:2017jye}, we can safely neglect stochastic sources after inflation because their contribution will be irrelevant compared to that of initial conditions. This is true due to the enhancement factor $e^{16|\xi|N_I}/\xi^2$, which is present in~(\ref{final_deSitter}) but not in the stochastic sources of  radiation epoch.

In radiation epoch ($\epsilon=2$) the Ricci scalar is null and we can also neglect $(m/H(t))^2$ (due to the assumption of very light scalar field, this term only becomes relevant at recent cosmological times, when matter and cosmological constant dominate the evolution of the Universe). This leads to a simplified evolution during radiation era, of the form

\begin{equation}
\frac{{\mathrm d}}{{\mathrm d} N}\Delta_{(\rm 4)}(N,r)+A~\Delta_{(\rm 4)}(N,r)=0\,,
\label{system_radiation}
\end{equation}
where A is the constant matrix
\begin{equation}
A=
\begin{bmatrix}
   0 & -1 & 0 & 0 & 0 & 0 \\
 0 & 1 & -2 & -2 & 0 & 0 \\
 0 & 0 & 2 & 0 & -2 & 0 \\
 0 & 0 & 0 & 2 & -1 & 0 \\
 0 & 0 & 0 & 0 & 3 & -4 \\
 0 & 0 & 0 & 0 & 0 & 4 \\
\end{bmatrix}\,.
\end{equation}
If $N=0$ is the beginning of radiation epoch, then the system is easily solved as $\Delta_{(\rm 4)}(N,r)=e^{-A~N}\Delta_{(\rm 4)}(0,r)$, where $\Delta_{(\rm 4)}(0,r)$ is the initial condition taken from the end of inflation~(\ref{final_deSitter}). The exponential of the matrix $-A$ is computed by diagonalizing $A$. At leading order in $|\xi|$ we find that, by the end of radiation epoch lasting $N_R\approx 50$ e-foldings,
\begin{equation}
\Delta_{(\rm 4)}(N_R,r)\simeq\frac{H_I^4 e^{16|\xi|N_I}}{1024\pi^4\xi^2}s(r)\left(1,16|\xi|e^{-N_R},64\xi^2e^{-2N_R},32\xi^2e^{-2N_R},256|\xi|^3e^{-3N_R},256\xi^4e^{-4N_R}\right)\,.
\label{final_radiation_0}
\end{equation}
Since $N_R\gg1$, we can say that, up to contributions suppressed by powers of $e^{-N_R}$, at the end of radiation epoch
\begin{equation}
\Delta_{(\rm 4)}(N_R,r)\simeq\frac{H_I^4}{1024\pi^4\xi^2}e^{16|\xi|N_I} s(r)(1,0,0,0,0,0)\,.
\label{final_radiation}
\end{equation}
This is the initial condition that will be used in the final stage of cosmological evolution after matter-radiation equality. We remark that the spatial dependence is still that imprinted by inflation. This is a consequence of the smallness of stochastic sources after inflation, so that the post-inflationary evolution does not affect the form of $s(r)$.

\subsection{Matter + cosmological constant epoch}
After matter-radiation equality, non-relativistic matter becomes the dominant component of the Universe leading its expansion. We know from eq. (90) of~\cite{Glavan:2017jye} that in matter period, the 2-pt functions of the scalar field evolve in such a way to develop a dark energy component (more precisely they have a matter-like contribution and a CC-like contribution). We are interested here in 4-pt functions, that should be evolved numerically in a background containing the backreaction of the scalar field. However we can obtain a simplified analytical treatment by approximating the background evolution as an exact matter+CC Universe.

We set $N=0$ at matter-radiation equality and the current (today) time corresponds to $N_0=\ln{\left(\frac{\Omega_M}{\Omega_R}\right)} \simeq8.1$. As usual, we denote by $\Omega_M$ and $\Omega_\Lambda=1-\Omega_M$ the matter and cosmological constant fractions of energy density today. Then $\epsilon(N)$ needed in~(\ref{eomPi2,Pi2}) evolves in time as
\begin{equation}
\epsilon(N)=\frac32 \frac{1}{1+\frac{\Omega_\Lambda}{\Omega_M}e^{3(N-N_0)}}\,.
\end{equation}
It is convenient to trade $N$ for the variable $x\equiv\left[1+\frac{\Omega_\Lambda}{\Omega_M}e^{3(N-N_0)}\right]^{-1}$ so that $\epsilon(x)=(3/2) x$, and with simple algebra the quantity $M(t)/H(t)$ appearing in~(\ref{eomPi2,Pi2}) becomes, in the $x$ variable,
\begin{equation}
\left(\frac{M(x)}{H(x)}\right)^2=\left(\frac{m}{H_{\rm DE}}\right)^2(1-x)-12|\xi|\left(1-\frac34 x\right)\,,
\end{equation}
where $H_{\rm DE}$ was already defined right after~(\ref{inflation_efolds_relation}).
The initial time $N=0$ (matter-radiation equality) corresponds to $x_{\rm eq}=\left(1+\frac{\Omega_\Lambda}{\Omega_M}e^{-3N_0}\right)^{-1}$ while the current time $N_0$ gives $x_0=\Omega_M$.
Using $dx/dN=3x(x-1)$ the 4-pt functions obey a system of the form (again we can neglect stochastic sources)
\begin{equation}
\frac{{\mathrm d}}{{\mathrm d} x}\Delta_{(\rm 4)}(x,r)+B(x)~\Delta_{(\rm 4)}(x,r)=0\,,
\label{system_matter+CC}
\end{equation}
 In the equation above B(x) is the time-dependent matrix $B(x)=b_1(x)B_1+b_2(x)B_2+b_3(x)B_3$, where the time dependence is encoded in functions
\begin{equation}
b_1(x)=\frac{2-x}{2x(x-1)}\,, \qquad b_2(x)=-\frac{\left(\frac{m}{H_{\rm DE}}\right)^2}{3x}-|\xi|\frac{4-3x}{x(x-1)}\,, \qquad b_3(x)=\frac{1}{3x(x-1)}\,,
\end{equation}
while $B_1$, $B_2$ and $B_3$ are the following constant matrices
\begin{equation}
B_1={\rm diag}(0,1,2,2,3,4)\,,
\qquad
B_2=
\begin{bmatrix}
 0 & 0 & 0 & 0 & 0 & 0 \\
 4 & 0 & 0 & 0 & 0 & 0 \\
 0 & 2 & 0 & 0 & 0 & 0 \\
 0 & 1 & 0 & 0 & 0 & 0 \\
 0 & 0 & 2 & 2 & 0 & 0 \\
 0 & 0 & 0 & 0 & 1 & 0 \\
\end{bmatrix}\,,
\qquad
B_3=
\begin{bmatrix}
 0 & -1 & 0 & 0 & 0 & 0 \\
 0 & 0 & -2 & -2 & 0 & 0 \\
 0 & 0 & 0 & 0 & -2 & 0 \\
 0 & 0 & 0 & 0 & -1 & 0 \\
 0 & 0 & 0 & 0 & 0 & -4 \\
 0 & 0 & 0 & 0 & 0 & 0 \\
\end{bmatrix}\,.
\end{equation}

The exact solution of~(\ref{system_matter+CC}) is
\begin{equation}
\Delta_{(\rm 4)}(x,r)=T\exp\left[-\int_{x_{\rm eq}}^x dx'~B(x')\right]\Delta_{(\rm 4)}(x_{\rm eq},r)\,,
\end{equation}
where $T\exp$ denotes the time-ordered exponential, which is there because $B(x)$ matrices at different $x$'s do not commute. However we checked numerically that neglecting the time ordering gives accurate results at a few percent level, with the advantage of displaying simpler formulas.\footnote{More details on how the exact evolution of correlators in matter+CC period can be computed will be given in \cite{inprep:2021etc}.} Then the integrals can be solved analytically and the matrix exponential can be also evaluated analytically with a symbolic computation software. Further simplifications arise by the assumptions of small $|\xi|$ and $\left(\frac{m}{H_{\rm DE}}\right)^2$, leading after a few steps to the result~(\ref{4pt_final}) at the current epoch $N_0$.

%BIBLIOGRAPHY


\begin{thebibliography}{99}

%\cite{Planck:2018vyg}
\bibitem{Planck:2018vyg}
N.~Aghanim \textit{et al.} [Planck],
``Planck 2018 results. VI. Cosmological parameters,''
Astron. Astrophys. \textbf{641} (2020), A6
[erratum: Astron. Astrophys. \textbf{652} (2021), C4]
%doi:10.1051/0004-6361/201833910
%[arXiv:1807.06209 [astro-ph.CO]].
%6090 citations counted in INSPIRE as of 20 Oct 2021

%\cite{Addison:2017fdm}
\bibitem{Addison:2017fdm}
G.~E.~Addison, D.~J.~Watts, C.~L.~Bennett, M.~Halpern, G.~Hinshaw and J.~L.~Weiland,
``Elucidating $\Lambda$CDM: Impact of Baryon Acoustic Oscillation Measurements on the Hubble Constant Discrepancy,''
Astrophys. J. \textbf{853} (2018) no.2, 119
%doi:10.3847/1538-4357/aaa1ed
%[arXiv:1707.06547 [astro-ph.CO]].
%156 citations counted in INSPIRE as of 20 Oct 2021

%\cite{Schoneberg:2019wmt}
\bibitem{Schoneberg:2019wmt}
N.~Sch\"oneberg, J.~Lesgourgues and D.~C.~Hooper,
``The BAO+BBN take on the Hubble tension,''
JCAP \textbf{10} (2019), 029
%doi:10.1088/1475-7516/2019/10/029
%[arXiv:1907.11594 [astro-ph.CO]].
%85 citations counted in INSPIRE as of 20 Oct 2021

%\cite{Jimenez:2001gg}
\bibitem{Jimenez:2001gg}
R.~Jimenez and A.~Loeb,
``Constraining cosmological parameters based on relative galaxy ages,''
Astrophys. J. \textbf{573} (2002), 37-42
%doi:10.1086/340549
%[arXiv:astro-ph/0106145 [astro-ph]].
%479 citations counted in INSPIRE as of 20 Oct 2021

%\cite{Haridasu:2018gqm}
\bibitem{Haridasu:2018gqm}
B.~S.~Haridasu, V.~V.~Lukovi\'c, M.~Moresco and N.~Vittorio,
``An improved model-independent assessment of the late-time cosmic expansion,''
JCAP \textbf{10} (2018), 015
%doi:10.1088/1475-7516/2018/10/015
%[arXiv:1805.03595 [astro-ph.CO]].
%70 citations counted in INSPIRE as of 20 Oct 2021

%\cite{Riess:2019cxk}
\bibitem{Riess:2019cxk}
A.~G.~Riess, S.~Casertano, W.~Yuan, L.~M.~Macri and D.~Scolnic,
``Large Magellanic Cloud Cepheid Standards Provide a 1\% Foundation for the Determination of the Hubble Constant and Stronger Evidence for Physics beyond $\Lambda$CDM,''
Astrophys. J. \textbf{876} (2019) no.1, 85
%doi:10.3847/1538-4357/ab1422
%[arXiv:1903.07603 [astro-ph.CO]].
%1043 citations counted in INSPIRE as of 20 Oct 2021

%\cite{Wong:2019kwg}
\bibitem{Wong:2019kwg}
K.~C.~Wong, S.~H.~Suyu, G.~C.~F.~Chen, C.~E.~Rusu, M.~Millon, D.~Sluse, V.~Bonvin, C.~D.~Fassnacht, S.~Taubenberger and M.~W.~Auger, \textit{et al.}
``H0LiCOW \textendash{} XIII. A 2.4 per cent measurement of H0 from lensed quasars: 5.3\ensuremath{\sigma} tension between early- and late-Universe probes,''
Mon. Not. Roy. Astron. Soc. \textbf{498} (2020) no.1, 1420-1439
%doi:10.1093/mnras/stz3094
%[arXiv:1907.04869 [astro-ph.CO]].
%498 citations counted in INSPIRE as of 20 Oct 2021

%\cite{Freedman:2020dne}
\bibitem{Freedman:2020dne}
W.~L.~Freedman, B.~F.~Madore, T.~Hoyt, I.~S.~Jang, R.~Beaton, M.~G.~Lee, A.~Monson, J.~Neeley and J.~Rich,
``Calibration of the Tip of the Red Giant Branch (TRGB),'' The Astrophysical Journal 891 (1) (2020) 57.
%doi:10.3847/1538-4357/ab7339
%[arXiv:2002.01550 [astro-ph.GA]].
%130 citations counted in INSPIRE as of 20 Oct 2021

%\cite{Huang:2018dbn}
\bibitem{Huang:2018dbn}
C.~D.~Huang, A.~G.~Riess, S.~L.~Hoffmann, C.~Klein, J.~Bloom, W.~Yuan, L.~M.~Macri, D.~O.~Jones, P.~A.~Whitelock and S.~Casertano, \textit{et al.}
``A Near-infrared Period\textendash{}Luminosity Relation for Miras in NGC 4258, an Anchor for a New Distance Ladder,''
Astrophys. J. \textbf{857} (2018) no.1, 67
%doi:10.3847/1538-4357/aab6b3
%[arXiv:1801.02711 [astro-ph.CO]].
%21 citations counted in INSPIRE as of 20 Oct 2021

%\cite{Blakeslee:2021rqi}
\bibitem{Blakeslee:2021rqi}
J.~P.~Blakeslee, J.~B.~Jensen, C.~P.~Ma, P.~A.~Milne and J.~E.~Greene,
``The Hubble Constant from Infrared Surface Brightness Fluctuation Distances,''
Astrophys. J. \textbf{911} (2021) no.1, 65
%doi:10.3847/1538-4357/abe86a
%[arXiv:2101.02221 [astro-ph.CO]].
%22 citations counted in INSPIRE as of 20 Oct 2021

%\cite{Pesce:2020xfe}
\bibitem{Pesce:2020xfe}
D.~W.~Pesce, J.~A.~Braatz, M.~J.~Reid, A.~G.~Riess, D.~Scolnic, J.~J.~Condon, F.~Gao, C.~Henkel, C.~M.~V.~Impellizzeri and C.~Y.~Kuo, \textit{et al.}
``The Megamaser Cosmology Project. XIII. Combined Hubble constant constraints,''
Astrophys. J. Lett. \textbf{891} (2020) no.1, L1
%doi:10.3847/2041-8213/ab75f0
%[arXiv:2001.09213 [astro-ph.CO]].
%119 citations counted in INSPIRE as of 20 Oct 2021

%\cite{DiValentino:2016hlg}
\bibitem{DiValentino:2016hlg}
E.~Di Valentino, A.~Melchiorri and J.~Silk,
``Reconciling Planck with the local value of $H_0$ in extended parameter space,''
Phys. Lett. B \textbf{761} (2016), 242-246
%doi:10.1016/j.physletb.2016.08.043
%[arXiv:1606.00634 [astro-ph.CO]].
%242 citations counted in INSPIRE as of 20 Oct 2021

%\cite{DiValentino:2019ffd}
\bibitem{DiValentino:2019ffd}
E.~Di Valentino, A.~Melchiorri, O.~Mena and S.~Vagnozzi,
``Interacting dark energy in the early 2020s: A promising solution to the $H_0$ and cosmic shear tensions,''
Phys. Dark Univ. \textbf{30} (2020), 100666
%doi:10.1016/j.dark.2020.100666
%[arXiv:1908.04281 [astro-ph.CO]].
%158 citations counted in INSPIRE as of 20 Oct 2021

%\cite{Li:2019san}
\bibitem{Li:2019san}
X.~Li, A.~Shafieloo, V.~Sahni and A.~A.~Starobinsky,
``Revisiting Metastable Dark Energy and Tensions in the Estimation of Cosmological Parameters,''
Astrophys. J. \textbf{887} (2019), 153
%doi:10.3847/1538-4357/ab535d
%[arXiv:1904.03790 [astro-ph.CO]].
%25 citations counted in INSPIRE as of 20 Oct 2021

%\cite{Vattis:2019efj}
\bibitem{Vattis:2019efj}
K.~Vattis, S.~M.~Koushiappas and A.~Loeb,
``Dark matter decaying in the late Universe can relieve the H0 tension,''
Phys. Rev. D \textbf{99} (2019) no.12, 121302
%doi:10.1103/PhysRevD.99.121302
%[arXiv:1903.06220 [astro-ph.CO]].
%123 citations counted in INSPIRE as of 20 Oct 2021

%\cite{Feeney:2018mkj}
\bibitem{Feeney:2018mkj}
S.~M.~Feeney, H.~V.~Peiris, A.~R.~Williamson, S.~M.~Nissanke, D.~J.~Mortlock, J.~Alsing and D.~Scolnic,
``Prospects for resolving the Hubble constant tension with standard sirens,''
Phys. Rev. Lett. \textbf{122} (2019) no.6, 061105
%doi:10.1103/PhysRevLett.122.061105
%[arXiv:1802.03404 [astro-ph.CO]].
%141 citations counted in INSPIRE as of 20 Oct 2021

%\cite{Lemos:2018smw}
\bibitem{Lemos:2018smw}
P.~Lemos, E.~Lee, G.~Efstathiou and S.~Gratton,
``Model independent $H(z)$ reconstruction using the cosmic inverse distance ladder,''
Mon. Not. Roy. Astron. Soc. \textbf{483} (2019) no.4, 4803-4810
%doi:10.1093/mnras/sty3082
%[arXiv:1806.06781 [astro-ph.CO]].
%63 citations counted in INSPIRE as of 20 Oct 2021

%\cite{Karwal:2016vyq}
\bibitem{Karwal:2016vyq}
T.~Karwal and M.~Kamionkowski,
``Dark energy at early times, the Hubble parameter, and the string axiverse,''
Phys. Rev. D \textbf{94} (2016) no.10, 103523
%doi:10.1103/PhysRevD.94.103523
%[arXiv:1608.01309 [astro-ph.CO]].
%154 citations counted in INSPIRE as of 20 Oct 2021

%\cite{Poulin:2018cxd}
\bibitem{Poulin:2018cxd}
V.~Poulin, T.~L.~Smith, T.~Karwal and M.~Kamionkowski,
``Early Dark Energy Can Resolve The Hubble Tension,''
Phys. Rev. Lett. \textbf{122} (2019) no.22, 221301
%doi:10.1103/PhysRevLett.122.221301
%[arXiv:1811.04083 [astro-ph.CO]].
%346 citations counted in INSPIRE as of 20 Oct 2021

%\cite{Ivanov:2020ril}
\bibitem{Ivanov:2020ril}
M.~M.~Ivanov, E.~McDonough, J.~C.~Hill, M.~Simonovi\'c, M.~W.~Toomey, S.~Alexander and M.~Zaldarriaga,
``Constraining Early Dark Energy with Large-Scale Structure,''
Phys. Rev. D \textbf{102} (2020) no.10, 103502
%doi:10.1103/PhysRevD.102.103502
%[arXiv:2006.11235 [astro-ph.CO]].
%76 citations counted in INSPIRE as of 20 Oct 2021

%\cite{Jedamzik:2020krr}
\bibitem{Jedamzik:2020krr}
K.~Jedamzik and L.~Pogosian,
``Relieving the Hubble tension with primordial magnetic fields,''
Phys. Rev. Lett. \textbf{125} (2020) no.18, 181302
%doi:10.1103/PhysRevLett.125.181302
%[arXiv:2004.09487 [astro-ph.CO]].
%61 citations counted in INSPIRE as of 20 Oct 2021

%\cite{Cai:2021wgv}
\bibitem{Cai:2021wgv}
R.~G.~Cai, Z.~K.~Guo, L.~Li, S.~J.~Wang and W.~W.~Yu,
``Chameleon dark energy can resolve the Hubble tension,''
Phys. Rev. D \textbf{103} (2021) no.12, 121302
%doi:10.1103/PhysRevD.103.L121302
%[arXiv:2102.02020 [astro-ph.CO]].
%10 citations counted in INSPIRE as of 20 Oct 2021

%\cite{Marra:2021fvf}
\bibitem{Marra:2021fvf}
V.~Marra and L.~Perivolaropoulos,
``Rapid transition of Geff at zt\ensuremath{\simeq}0.01 as a possible solution of the Hubble and growth tensions,''
Phys. Rev. D \textbf{104} (2021) no.2, L021303
%doi:10.1103/PhysRevD.104.L021303
%[arXiv:2102.06012 [astro-ph.CO]].
%19 citations counted in INSPIRE as of 20 Oct 2021

%\cite{Glavan:2017jye}
\bibitem{Glavan:2017jye}
D.~Glavan, T.~Prokopec and A.~A.~Starobinsky,
``Stochastic dark energy from inflationary quantum fluctuations,''
Eur. Phys. J. C \textbf{78} (2018) no.5, 371
%doi:10.1140/epjc/s10052-018-5862-5
%[arXiv:1710.07824 [astro-ph.CO]].
%21 citations counted in INSPIRE as of 12 Oct 2021

%\cite{Glavan:2015cut}
\bibitem{Glavan:2015cut}
D.~Glavan, T.~Prokopec and T.~Takahashi,
``Late-time quantum backreaction of a very light nonminimally coupled scalar,''
Phys. Rev. D \textbf{94} (2016), 084053
%doi:10.1103/PhysRevD.94.084053
%[arXiv:1512.05329 [gr-qc]].
%15 citations counted in INSPIRE as of 12 Nov 2021

%\cite{Glavan:2014uga}
\bibitem{Glavan:2014uga}
D.~Glavan, T.~Prokopec and D.~C.~van der Woude,
``Late-time quantum backreaction from inflationary fluctuations of a nonminimally coupled massless scalar,''
Phys. Rev. D \textbf{91} (2015) no.2, 024014
%doi:10.1103/PhysRevD.91.024014
%[arXiv:1408.4705 [gr-qc]].
%22 citations counted in INSPIRE as of 12 Nov 2021

%\cite{Glavan:2013mra}
\bibitem{Glavan:2013mra}
D.~Glavan, T.~Prokopec and V.~Prymidis,
``Backreaction of a massless minimally coupled scalar field from inflationary quantum fluctuations,''
Phys. Rev. D \textbf{89} (2014) no.2, 024024
%doi:10.1103/PhysRevD.89.024024
%[arXiv:1308.5954 [gr-qc]].
%30 citations counted in INSPIRE as of 12 Nov 2021

%\cite{inprep:2021etc}
\bibitem{inprep:2021etc}
E.~Belgacem, T.~Prokopec,
``Spatial correlations of dark energy from quantum fluctuations in inflation,''
in preparation.

%\cite{Demianski:2019vmq}
\bibitem{Demianski:2019vmq}
M.~Demianski and E.~Piedipalumbo,
``Observational tests of the Glavan, Prokopec and Starobinsky model of dark energy,''
Eur. Phys. J. C \textbf{79} (2019) no.7, 575
%doi:10.1140/epjc/s10052-019-7045-4
%[arXiv:1906.06107 [astro-ph.CO]].

%\cite{Nan:2021prt}
\bibitem{Nan:2021prt}
Y.~Nan and K.~Yamamoto,
``Dark energy model with very large-scale inhomogeneity,''
[arXiv:2111.14174 [astro-ph.CO]].

\bibitem{Troster:2019ean}
Tr\"oster, Tilman and others,
``Cosmology from large-scale structure: Constraining $\Lambda$CDM with BOSS",
[arxiv:1909.11006],
Astron. Astrophys.633, L10, 2020


\end{thebibliography}
\end{document}